\documentclass[prb,aps,twocolumn,superscriptaddress,longbibliography]{revtex4-2}

\usepackage{amsmath,amssymb}
\usepackage{graphicx}
\usepackage{bm}
\usepackage{hyperref}

\graphicspath{{figures/}}
\begin{document}

\title{Theoretical Prediction of Optimal $T_c$ and Fermi Pockets in Nickelate Superconductors}
\author{Xiuqing Huang}
\email{xiuqing\_huang@163.com}

\affiliation{Department of Telecommunications Engineering ICE, Army Engineering University of PLA, Nanjing 210007, China}
\affiliation{National Laboratory of Solid State Microstructures of Physics, Nanjing
	University, Nanjing 210093, China}

\begin{abstract}
	High-pressure bilayer $\mathrm{La_{3-x}Sm_{x}Ni_{2}O_{7-\delta}}$ (LSNO) reaches a record $T_\mathrm{c}=96\ \mathrm{K}$, triggering wide discussion on the $T_\mathrm{c}$ ceiling of nickelate superconductors. We show monoclinic and tetragonal LSNO share the same octahedral quantum-well motif governing $T_\mathrm{c}$ with $\mathrm{YBa_{2}Cu_{3}O_{7-\delta}}$ (YBCO). Using the Planckian quantum-well scaling $T_\mathrm{c} = \Lambda/\xi^{2}$ ($\xi$: lattice-modulated quantum-well depth), we obtain $T_\mathrm{c}=93.4\ \mathrm{K}$ and $97.1\ \mathrm{K}$ for monoclinic and tetragonal LSNO, matching experimental values $92\ \mathrm{K}$ and $96\ \mathrm{K}$. Despite distinct stoichiometry and global symmetry ($P2_1/m$ for LSNO, $Pmmm$ for orthorhombic YBCO), both systems have nearly identical $\xi$ ($3.6629\ \mathrm{\AA}$ vs. $3.6720\ \mathrm{\AA}$) and consistent $T_\mathrm{c}$ responses. Further calculations yield a universal $T_\mathrm{c}$ limit $\sim100\ \mathrm{K}$ for rare-earth nickelates, irrespective of stacking sequences. 	
	We examine four nickelate multilayer stacking variants: 2222 (pure bilayer), 1212 (alternating single-bilayer), 2323 (bilayer–trilayer), and 1313 (single-trilayer). Mirror symmetry breaking of coupled twin quantum wells, unique to bilayer nickelates, dictates $\gamma$ Fermi pocket formation and ambient-pressure superconductivity. We further prove Fermi surfaces constitute a hologram of quantum-well electrons, establishing intrinsic links between quantum-well symmetry breaking, Fermi pocket structural evolution, and superconducting properties.
	
\end{abstract}

\maketitle

\section{Introduction}

The discovery of superconductivity at $T_\mathrm{c}\sim80\ \mathrm{K}$ in high-pressure bilayer Ruddlesden--Popper (R--P) $\mathrm{La_3Ni_2O_7}$ single crystals \cite{sun2023signatures} has triggered intensive experimental \cite{chengjg_nature,shi2025spin,Li2025NSRev,Zhang2024,chen2024electronic,chen2024evidence,zhu2024superconductivity,puphal2024unconventional,ranna2025disorder,zhang2025superconductivity,ko2025signatures,zhou2025ambient,sahib2025superconductivity,abadi2025electronic,pan2026phase,wen2026tunnel,li2026reentrant} and theoretical investigations \cite{qu2024bilayer,sakakibara2024possible,lu2024interlayer,jiang2024pressure,zhang2024strong,jiang2025theory,zhan2025cooperation,chow2025bulk,luo2023bilayer}. Despite considerable progress in synthesizing various superconducting nickelate variants through different strategies, their superconducting critical temperatures remain saturated near $80\ \mathrm{K}$ \cite{chengjg_nature,shi2025spin,Li2025NSRev,Zhang2024}. To date, no unified microscopic mechanism or quantitative $T_\mathrm{c}$-prediction theory has been established, leaving the exploration of high-$T_\mathrm{c}$ nickelate superconductors largely empirical. 
A substantial advance was achieved in ambient-pressure flux-grown Sm-doped nickelates, which elevates the maximum $T_\mathrm{c}$ from $83\ \mathrm{K}$ to $96\ \mathrm{K}$ \cite{Ni96K}. This breakthrough significantly promotes nickelate superconductivity research and revitalizes the prospect of realizing cuprate-compatible high-temperature superconductivity. Most recently, high-pressure Nd-substituted $\mathrm{La_{3-x}Nd_{x}Ni_{2}O_{7}}$ bilayer nickelates exhibit local superconducting signatures up to $100.5\ \mathrm{K}$ via radio-frequency detection, though full resistive zero crossing is absent in the bulk sample \cite{Qiu2026La3xNdxNi2O7}.

Complementary progress has been made via the gigantic oxidative atomic-layer-by-layer epitaxy (GAE) technique, which enables the fabrication of four artificially structured nickelate superstructures with programmable atomic stacking sequences: pure bilayer (2222), 1212 single-bilayer, 1313 single-trilayer, and 2323 bilayer-trilayer configurations \cite{xue2026superstructure}. At ambient pressure, the 1212, 2222, and 2323 phases exhibit superconductivity exceeding the conventional McMillan limit, while the 1313 phase remains non-superconducting. Angle-resolved photoemission spectroscopy measurements reveal emergent $\gamma$-band Fermi pockets near the Brillouin zone corners in superconducting films, which are absent in the non-superconducting 1313 counterpart. The strong stacking dependence of superconductivity, especially the drastically distinct superconducting behaviors between the 2222 and 1313 phases with identical chemical composition, suggests that superconductivity may be governed by certain local structural units, posing a severe challenge to existing superconducting theories and pairing mechanisms \cite{shen2026nodeless,chen2026pairing1313}.

Recent years have witnessed extensive reviews and rapid progress in nickel-based superconductivity, ushering in the ``nickelate era'' of superconducting research \cite{112-review-1,112-review-2,review-cpl,jiang2025nsrreview,nickel-age}. Nonetheless, most conventional theories originating from cuprate and iron-based superconductors lack profound physical implications and frequently lead to inconsistent predictions.
For bilayer nickelate $\mathrm{R_3Ni_2O_7}$ ($\mathrm{R}=\text{La}$--$\text{Sm}$), contradictory $T_c$ evolution scenarios have been proposed: $T_c$ may decrease with expanding rare-earth ionic radii \cite{zhang2023trends}, or alternatively rise toward room temperature \cite{Pan2024Effect,Geisler2024}.
Traditional theoretical models often generate superficial or misleading results for mainstream superconducting systems. Tiny variations in temperature, pressure, doping, or atomic stacking can trigger complex phase transitions that cannot be quantitatively modeled. In the typical $\text{KFeSe}$ system, pressure above $10.5\:$GPa drastically elevates $T_c$ from $0 ~$K to over $40~$K \cite{sun2012reemerging}. This unclarified anomalous behavior strongly demands a novel fundamental paradigm for superconductivity studies.

Conventional superconductivity, as a lossless coherent condensate, is positively correlated with metallicity and lattice perfection. In contrast, high-$T_\text{c}$ superconductors are governed by dissipative effects such as insulating behavior, disorder, and doping. Recent experiments confirm $T_\text{c}\propto\sqrt{A_1}$, where $A_1$ denotes the linear resistance coefficient \cite{Yuan2022}. High-$T_\text{c}$ cuprate, iron-based, and nickel-based superconductors all require parent-phase doping, which contradicts the conventional free-electron superconducting picture. To resolve this inconsistency, we attribute superconductivity to the Planck quantum ground state and establish the quantum-well-localized electron and electron–hole symmetry-breaking pairing (QSP) theory \cite{huang1}, which well explains and predicts key experimental results of cuprate and iron-based systems \cite{huang2,huang3}.

This work focuses on the structural-based optimal $T_\text{c}$ prediction for unconventional superconductors, particularly newly discovered nickelates \cite{Ni96K,xue2026superstructure}. We extend the quantum-well mechanism to four nickelate superstructures (2222, 1212, 1313, 2323) and reveal the inherent correlation between Fermi pocket topology, quantum-well symmetry breaking and stacking configurations, providing new insights into the unified mechanism of unconventional superconductivity. The remainder of this paper is organized as follows. Sec.\ref{sec:Tc} presents the unified experimental-validated $T_\text{c}$ formula based on quantum-well structures of cuprate and iron-based superconductors. Sec.\ref{sec:optimal} analyzes nickelate quantum-well symmetry breaking and predicts their optimal $T_\text{c}$. Sec.\ref{sec:1313} clarifies the interplay among symmetry breaking, Fermi surface features and superconductivity for four nickelate phases. Sec.\ref{sec:conclusion} concludes the main results and prospects future studies.

\section{Planckian Quantum Well and Unified $T_\text{c}$ Formula}\label{sec:Tc}

Over more than a century of superconducting research, and particularly since the discovery of layered high-$T_\text{c}$ superconductors, abundant experimental evidence has consistently demonstrated a pronounced correlation between the superconducting critical temperature $T_\text{c}$ and the $c$-axis lattice parameter, known as the $c$-axis effect \cite{leggett1999}. Applying pressure, chemical substitution, and structural intercalation can effectively compress the interlayer spacing along the $c$ axis and thereby raise $T_\text{c}$. Quantitatively establishing the relationship between the $c$-axis effect and $T_\text{c}$ constitutes a crucial route toward unraveling the mystery of high-temperature superconductivity. In this work, we show that the $c$-axis effect encodes a more fundamental crystalline structural unit: the Planck quantum well, where the localized electrons are Planck quantum harmonic oscillators.  This unit uniquely determines the ground-state energy within the Planck quantum framework and simultaneously governs the optimal superconducting transition temperature as well as numerous key experimental observations of superconductivity.

\begin{figure}[t]
	\centering 
	\includegraphics[width=1.0\columnwidth]{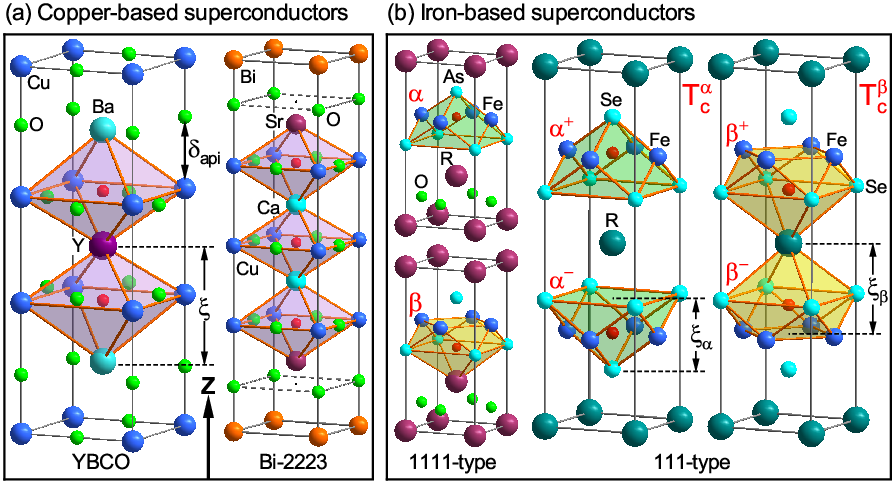} 
	\caption{Planckian Quantum Wells and Localized Electrons in Unconventional Superconductors. 		
		(a) Octahedral quantum wells in cuprates (YBCO, Bi-2223): their geometry and density dictate Fermi surface structure and symmetry, with well depth $\xi$ governing optimal $T_{c}$. 		
		(b) Polyhedral quantum wells of 1111/111-type iron-based superconductors ($R$: substitutable elements). Unlike cuprates, they possess two distinct well types ($\alpha$, $\beta$) with depths $\xi_{\alpha}$, $\xi_{\beta}$ and corresponding dual superconducting phases. Lattice mirror symmetry ensures paired upward/downward quantum wells ($\alpha^{\pm}$, $\beta^{\pm}$). Since $\xi_{\alpha}<\xi_{\beta}$, hence $T_{c}^{\alpha}>T_{c}^{\beta}$. Pressure reduces $\xi_{\beta}$, driving originally localized $\beta$-pairing electrons in the Se plane into the high-$T_{c}$ $\alpha$-Fe plane via enhanced Coulomb repulsion—this is the microscopic mechanism of dual-superconducting-phase transition in iron-based superconductors \cite{sun2012reemerging}.}
	\label{figure1}
\end{figure}

As shown in Fig. \ref{figure1}(a), cuprate superconductors possess an octahedral quantum well structure (purple) with electrons localized in Cu--O planes. Fig. \ref{figure1}(b) depicts two iron-based superconductor families comprising two quantum well types: green $\alpha$-type tent-shaped nonahedra (with electrons confined to Fe layers) and yellow $\beta$-type diamond-structured tridecahedra (localized in As/Se layers). It is noteworthy that identically configured polyhedrons exist as oppositely oriented pairs. These nanoscale polyhedral clusters (quantum wells), functioning as superconducting ``DNA", govern key electronic and structural properties (e.g., Fermi surface topology, pseudogap formation, charge order phenomena, and pressure/doping-dependent phase transitions). Iron-based superconductors feature more complex quantum well geometries than cuprates, giving rise to a richer spectrum of emergent physical phenomena \cite{iron-review1,iron-review2,iron-review3}.

Of particular significance, the best-performing conventional conductors (Au, Ag, Cu) are non-superconducting, whereas insulating ceramics can exhibit high-temperature superconductivity. A long-overlooked critical difference lies in their mechanical ductility: Au, Ag, and Cu are highly ductile, while ceramics are extremely brittle. This raises a fundamental question: is there a correlation between superconductivity and brittleness, such that the brittler the material, the higher the achievable $T_{\rm c}$? Brittleness essentially originates from the localization of valence electrons. Cooling, pressurization, chemical doping, and ionic substitution enhance electron localization and material brittleness, thereby increasing the threshold of electron thermal vibrations (i.e., the intrinsic resonance frequency $\omega$). The quantum wells in Fig. \ref{figure1} act as Planck resonators. According to the law of energy conservation, the thermal energy (potential energy) at the superconducting critical temperature is equal to the Planck ground-state vibrational energy (kinetic energy): $k_{\rm B}T_{\rm c} = \hbar\omega$. In this case, Planck quantum radiation is completely suppressed, and the material enters a lossless superconducting state. Therefore, enhancing brittleness or increasing $\omega$ is the key to raising $T_{\rm c}$. It is not difficult to deduce that the crystal-structure-determined quantum well depth $\xi$ (including $\xi_\alpha$ and $\xi_\beta$ for iron-based systems) and the optimal $T_{\rm c}$ satisfy the following unified relationship:

\begin{equation}
	T_{c} = \frac{2e\hbar}{\pi k_{B}} \sqrt{\frac{\sigma}{m_{e}\varepsilon_{0}}} \frac{1}{\xi^{2}} = \frac{\Lambda}{\xi^{2}},
	\label{Tc}
\end{equation}
where $e$ is the electron elementary charge, $\hbar$ the reduced Planck constant, $k_{B}$ the Boltzmann constant, $m_{e}$ the electron rest mass, $\varepsilon_{0}$ the vacuum permittivity, $\sigma$ an undetermined fitting constant correlated with superconductor brittleness, and $\Lambda$ the electron-quantum well coupling strength, which correlates with the electron-confining plane's electronic and structural properties.

\begin{table}
	\caption{Relationship between $T_{c}$ and localized electron quantum well depth $\xi$ in Cu- and Fe-based superconductors.
		1st column: Superconductor compounds. 2nd column: $T_{c}$. 3rd–5th columns: Quantum well depths (bold = $T_{c}$-corresponding superconducting phases). 6th column: Correlation strength factor ($\Lambda=T_{c}\xi^{2}$).}

	\global\long\def\arraystretch{1}%
	\centering %
	\begin{tabular}{cccccc}
		\textbf{Compound} & $\textbf{T}_{\textbf{c}}$(K) & $\xi_{\alpha}$ (\AA) & $\xi_{\beta}$ (\AA) & $\xi_{\gamma}$ (\AA) & $\Lambda$\tabularnewline
		\hline	\hline 
		$\mathrm{YBa_{2}Cu_{3}O_{7-\delta}}$ & 93 & ${\boldsymbol{3.6720}}$ &  &  & 1254\tabularnewline
		$\mathrm{Bi_{2}Sr_{2}Ca_{2}Cu_{3}O_{10+\delta}}$ & 110 & 3.1511 & ${\boldsymbol{3.3272}}$ &  & 1217\tabularnewline
		$\mathrm{TlBa_{2}CaCu_{2}O_{7+\delta}}$ & 103 & ${\boldsymbol{3.5771}}$ &  &  & 1317\tabularnewline
		$\mathrm{TlBa_{2}Ca_{3}Cu_{4}O_{11+\delta}}$ & 112 & 3.2453 & ${\boldsymbol{3.5025}}$ &  & 1373\tabularnewline
		$\mathrm{TlBa_{2}Ca_{2}Cu_{3}O_{10+\delta}}$ & 120 & ${\boldsymbol{3.2381}}$ & 3.5305 &  & 1258\tabularnewline
		$\mathrm{Tl_{2}Ba_{2}Ca_{2}Cu_{3}O_{10+\delta}}$ & 128 & ${\boldsymbol{3.3053}}$ & 3.5252 &  & 1393\tabularnewline
		$\mathrm{HgBa_{2}CuO_{4+\delta}}$ & 94 & ${\boldsymbol{3.8241}}$ &  & 1374\tabularnewline
		$\mathrm{HgBa_{2}Ca_{2}Cu_{3}O_{8+\delta}}$ & 134 & ${\boldsymbol{3.2108}}$ & 3.4241 &  & 1380\tabularnewline
		\hline\ 
		$\mathrm{LaFeAsO_{1-x}F_{x}}$ & 43 & 2.6723 & ${\boldsymbol{3.0749}}$ &  & 406\tabularnewline
		$\mathrm{SmFeAsO_{1-x}F_{x}}$ & 43 & 2.7291 & ${\boldsymbol{3.0792}}$ &  & 407\tabularnewline
		$\mathrm{SmFeAsO_{1-x}F_{x}}$ & 55 & ${\boldsymbol{2.7149}}$ & 3.0433 &  & 405\tabularnewline
		$\mathrm{TbFeAsO_{1-x}F_{x}}$ & 45 & 2.7567 & ${\boldsymbol{3.0501}}$ &  & 418\tabularnewline
		$\mathrm{GdFeAsO_{1-x}F_{x}}$ & 53.5 & ${\boldsymbol{2.7454}}$ & 3.0568 &  & 403\tabularnewline
		$\mathrm{PrFeAsO_{1-x}F_{x}}$ & 52 & ${\boldsymbol{2.6857}}$ & 3.0591 &  & 375\tabularnewline
		$\mathrm{NdFeAsO_{1-x}F_{x}}$ & 50 & ${\boldsymbol{2.7227}}$ & 3.0418 &  & 371\tabularnewline
		$\mathrm{CeFeAsO_{1-x}F_{x}}$ & 41 & 2.6930 & ${\boldsymbol{3.0777}}$ &  & 388\tabularnewline
		$\mathrm{LaFeAsO}$ & 41 & 2.6608 & ${\boldsymbol{3.0912}}$ &  & 392\tabularnewline
		$\mathrm{LaYFeAsO}$ & 42 & 2.6840 & ${\boldsymbol{3.0912}}$ &  & 401\tabularnewline
		$\mathrm{BaKFe_{2}As_{2}}$ & 38 & 2.7650 & ${\boldsymbol{3.3242}}$ &  & 419\tabularnewline
		$\mathrm{CaKFe_{4}As_{4}}$ & 35 & 2.5617 & 2.9276 & ${\boldsymbol{3.3874}}$ & 402\tabularnewline
		$\mathrm{KCa_{2}Fe_{4}As_{4}F_{2}}$ & 33 & 2.8406 & 3.0301 & ${\boldsymbol{3.4353}}$ & 390\tabularnewline
		$\mathrm{K_{0.8}Fe_{y}Se_{2}}$ & 32 &  & ${\boldsymbol{3.534}}$ &  & 400\tabularnewline
		& 48.7 & ${\boldsymbol{2.836}}$ &  &  & 393\tabularnewline
		\hline
	\end{tabular}\label{tab:Tc}
\end{table}

A key point is that the critical temperature $T_c$ of Eq.(\ref{Tc}) incorporates nearly all fundamental physical constants. In particular, if $\hbar$, then $T_c=0$ necessarily holds, confirming that superconductivity is a quintessential quantum effect. Table \ref{tab:Tc} shows the measured superconducting transition temperature $T_c$ of cuprate and iron-based superconductors versus polyhedral quantum well depth $\xi$, following the inverse-square law in Eq. (\ref{Tc}) with correlation strength factors $\Lambda(\mathrm{Cu})\simeq1300$ and $\Lambda(\mathrm{Fe})\simeq400$ (threefold difference, consistent with record $T_c$: 164 K for cuprates \cite{cu164K}, 55 K for iron-based \cite{Ren2008}). As shown in Fig. \ref{figure1}(b), $\xi_{\alpha}<\xi_{\beta}$ for all iron-based superconductors, hence $T_{c}^{\alpha}>T_{c}^{\beta}$, leading to two paired quantum well structures and two $T_c$ domes. The low-$T_c$ $\beta$-phase exists at ambient/low pressure, while the high-$T_c$ $\alpha$-phase requires high pressure; their transition arises from electron pair transfer from Fe to Se (As) layers via Coulomb repulsion \cite{huang1}, verified experimentally \cite{sun2012reemerging}. According to Table \ref{tab:Tc}, the double-dome phenomenon is also observable in cuprate superconductors \cite{cooper2009anomalous}.
For $\mathrm{SmFeAsOF}$, experimental $T_c$ is 43 K (ambient pressure \cite{chen2008}) and 55 K (high pressure \cite{Ren2008}), matching our predictions (42.2 K, 54.3 K). In high-pressure $\mathrm{KFeSe}$, experimental dual transitions (32 K, 48.7 K \cite{sun2012reemerging}) agree well with our predictions (32.1 K, 49 K). Quantum wells act as resonant cavities for Planck-quantized electrons; lattice symmetry enforces top-bottom pairing and resonance, with localized electron resonance energy $E_R\propto1/\xi^2$. For iron-based superconductors, $E_R\simeq115/\xi^2$ (meV). Neutron spin scattering on $\mathrm{CaKFe_4As_4}$ (CKFA) \cite{xie2018} shows peaks at 9.5, 13.0, 18.3 meV, consistent with our theoretical three resonant cavities (one $\alpha$-type, two $\beta$-types) with energies 9.9, 13.4, 17.6 meV. Within the same quantum-well framework, we recently find that the superconducting gaps from STS spectra also follow an inverse-square relation with the quantum-well depth $\xi$  For $\mathrm{KCa_2Fe_4As_4F_2}$, our predicted gap pairs of $\pm6.2$ meV, $\pm5.6$ meV, and $\pm4.2$ meV are in excellent agreement with experimental results of $\pm6.2$ meV, $\pm5.4$ meV, and $\pm4.4$ meV \cite{huang3}.

\section{Optimal $T_\text{c}$ of Nickelate Superconductors} \label{sec:optimal}

\begin{figure}[t]
	\includegraphics[width=1.0\columnwidth]{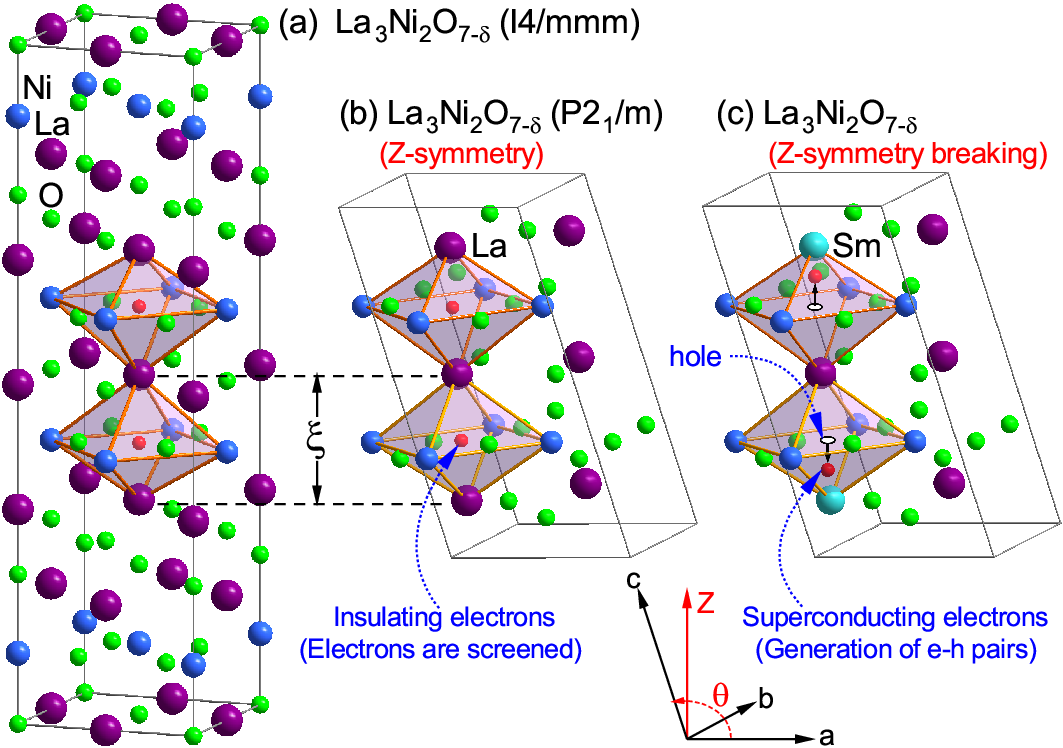} %\end{center}
	\caption{Octahedral Quantum Wells and Localized Electrons in Nickel-based Superconductors.
		(a) Tetragonal and orthorhombic bilayer nickel-based superconductors.
		(b) Undoped monoclinic nickel-based superconductors (monoclinic angle $\theta$): full quantum well symmetry screens localized electrons, resulting in an insulating state.
		(c) Ionic substitution breaks quantum well mirror symmetry along the $Z$-axis (octahedron vertex connecting direction), generating electron-hole dipoles and inducing the insulator-to-superconductor phase transition.}
	\label{figure2}
\end{figure}

Shortly after nickel-based superconductor discovery, we predicted their upper-limit critical temperature $T_c\approx100\ \text{K}$ from lattice structure \cite{huang1}. As shown in Fig. \ref{figure2}, nickelates exhibit octahedral quantum wells geometrically and structurally identical to cuprates (regardless of orthorhombic, tetragonal or monoclinic phases), with comparable correlation strength. For all reported nickelate systems (monolayer, bilayer, multilayer), quantum well depth $\xi=3.6$–$3.7\ \text{\AA}$, corresponding to optimal $T_c\approx100\ \text{K}$. Thus, Li \textit{et al.}'s $96\ \text{K}$ \cite{Ni96K} is close to the optimal value, as further verified below.

First, we resolve a long-standing superconductivity puzzle: why pressure, oxygen doping and ionic substitution induce superconducting transitions in insulating parents and modulate $T_c$? As shown in Fig. \ref{figure2}(a), tetragonal nickelates only need pressure for insulator-to-superconductor transition. In contrast, monoclinic nickelates require elemental substitution plus pressure to transition from insulation (Fig. \ref{figure2}(b)) to superconductivity (Fig. \ref{figure2}(c)). The key mechanism is quantum well symmetry breaking: octahedral quantum wells must break Z-axis mirror symmetry (vertex-connecting line) to form electron-hole dipoles, which interact with external electromagnetic fields and trigger superconducting transitions.

\begin{figure*}[t]
	\includegraphics[width=2\columnwidth]{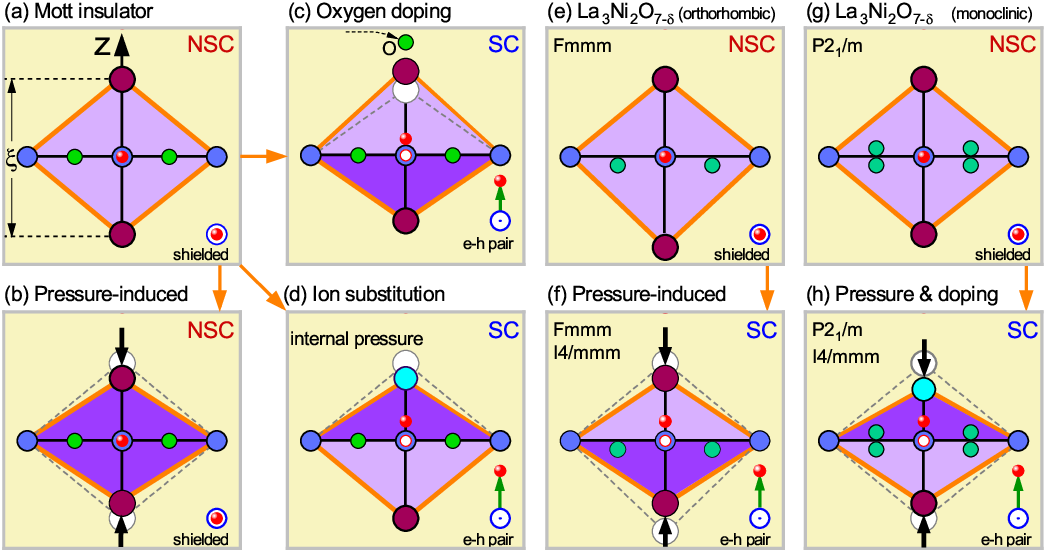} %\end{center}
	\caption{Mirror Symmetry Breaking and Insulator-to-Superconductor Transition Induced by Pressure, Oxygen Doping and Ionic Substitution (Internal Pressure)
		(a) Symmetric Mott insulator quantum well: localized electrons are symmetry-screened (inset, bottom right; electroneutrality makes octahedral quantum well equivalent to a hole), leading to insulating state (NSC).
		(b) Pressure preserves symmetry and insulation.
		(c) Oxygen doping breaks mirror symmetry (schematic, bottom right), generates electron-hole dipoles (e-h pairs), and triggers insulator-to-superconductor transition (SC).
		(d) Ionic substitution similarly induces symmetry breaking and superconductivity.
		(e)-(f) Intrinsically asymmetric orthorhombic $\mathrm{La_3Ni_2O_7}$ achieves superconductivity via pressure alone.
		(g)-(h) Intrinsically symmetric monoclinic $\mathrm{La_3Ni_2O_7}$ requires both ionic substitution and pressure for symmetry breaking and superconductivity.}
	\label{figure3}
\end{figure*}

Within the QSP framework, superconducting phase transition is the response of single-quantum-well localized electrons to external stimuli, due to superconducting electron indistinguishability. Fig. 3(a) shows high-$T_c$ superconductors' Mott insulating parent phase: full octahedral quantum well symmetry screens localized electrons, leading to insulation. As in Fig. \ref{figure3}(b), pressure alone reduces $\xi$ but preserves Z-axis mirror symmetry, maintaining insulation. In Fig. \ref{figure3}(c), oxygen doping at vertex vacancies breaks symmetry; Coulomb interactions drive electrons to new equilibria, forming electron-hole dipoles and inducing superconductivity. Similarly, ionic substitution at any octahedral vertex (Fig. \ref{figure3}(d)) breaks symmetry and triggers insulator-to-superconductor transition. Fig. \ref{figure3}(e) presents orthorhombic nickelate quantum well structure: non-coplanar O/Ni atoms induce intrinsic asymmetry, but electrical neutrality retains screening/insulation. Unlike Fig. \ref{figure3}(a), pressure here causes asymmetric effects (due to intrinsic upper-lower differences) to trigger superconductivity (Fig. \ref{figure3}(f)). For monoclinic structure (Fig. \ref{figure3}(g)), octahedral symmetry resembles Fig. \ref{figure3}(a); symmetry breaking and superconductivity require combined ionic substitution and pressure. Pressure magnitude induces two phases (large-$\xi$ monoclinic, small-$\xi$ tetragonal), corresponding to two $T_c$-distinct superconducting phases.

\begin{figure}[t]
	\centering
	\includegraphics[width=1.0\columnwidth]{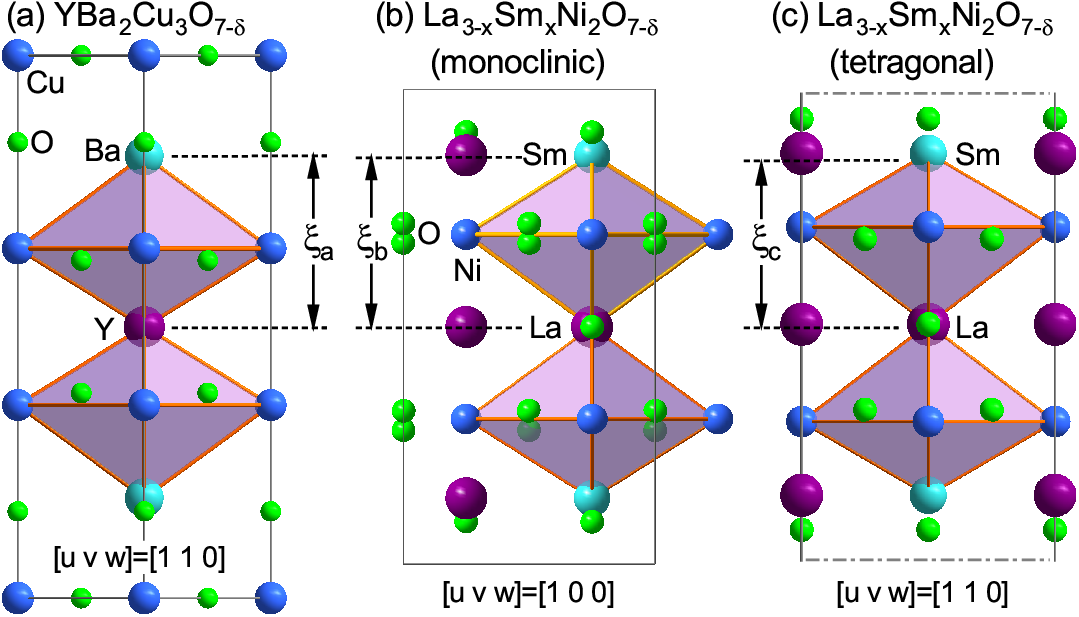}
	\caption{Comparison of Polyhedral Quantum Well Structures between Cuprate YBCO and Nickelate LSNO.
		(a) Projection of YBCO (Fig. \ref{figure1}(a)) along $[110]$ direction;
		(b)-(c) Projections of monoclinic LSNO (Fig. \ref{figure2}(c)) along $[100]$ and tetragonal LSNO (Fig. \ref{figure2}(a)) along $[110]$, respectively. They share nearly identical octahedral quantum well structures with $\xi_a \approx \xi_b > \xi_c$, indicating analogous superconducting origins. Based on localized quantum well superconductivity theory, their $T_c$ values are 93 K, 93.4 K and 97.1 K, in excellent agreement with experimental data (93 K, 92 K, 96 K).}
	\label{figure4}
\end{figure}

To further verify the reliability of the QSP theory, particularly the validity of Eq. (\ref{Tc}) in predicting the
superconducting transition temperature $T_{c}$, we obtained the crystal
structure experimental data of LSNO from the corresponding author of the original paper. For a
more reasonable comparison, we selected YBCO with uneven copper-oxygen planes (Fig. \ref{figure4}(a)) as the reference
material. It is striking to find that the octahedral structures of
the three compounds are extremely similar, regardless of whether they
are monoclinic (Fig. \ref{figure4}(b)) or tetragonal (Fig. \ref{figure4}(c)); specifically,
the structures in Fig. \ref{figure4}(a) and Fig. \ref{figure4}(c) are nearly identical. Based
on Eq. (\ref{Tc}), it can be preliminarily inferred that their $T_{c}$
values should be very close. For quantitative estimation, we directly
adopted the correlation strength factor $\Lambda=1254$
of YBCO with $\xi_{a}=3.6720\ \text{\AA}$
from Table \ref{tab:Tc}. For the monoclinic nickel-based superconductor in Fig.
4(b), the experimental value of the quantum well depth is $\xi_{b}=3.6629\ \text{\AA}$, and the $T_{c}$ value
determined by Eq. (\ref{Tc}) is approximately $93.4\ \text{K}$,
which is in excellent agreement with the experimental result of $92\ \text{K}$. The structural phase transition from
Fig. \ref{figure4}(b) to Fig. \ref{figure4}(c) induces a decrease in $\xi$
and an increase in $T_{c}$, where the increment can be estimated
by the following formula:

\begin{equation}
	\Delta T_{c}=-\frac{2\Lambda}{\xi^{3}}\Delta\xi = -2T_{c}\frac{\Delta\xi}{\xi},
	\label{dTc}
\end{equation}
where the negative sign indicates that $\Delta\xi < 0$  leads to $\Delta T_{c} > 0$, consistent with experimental $T_{c}$ enhancement upon lattice contraction.

\begin{figure}[t]
	\centering
	\includegraphics[width=1\columnwidth]{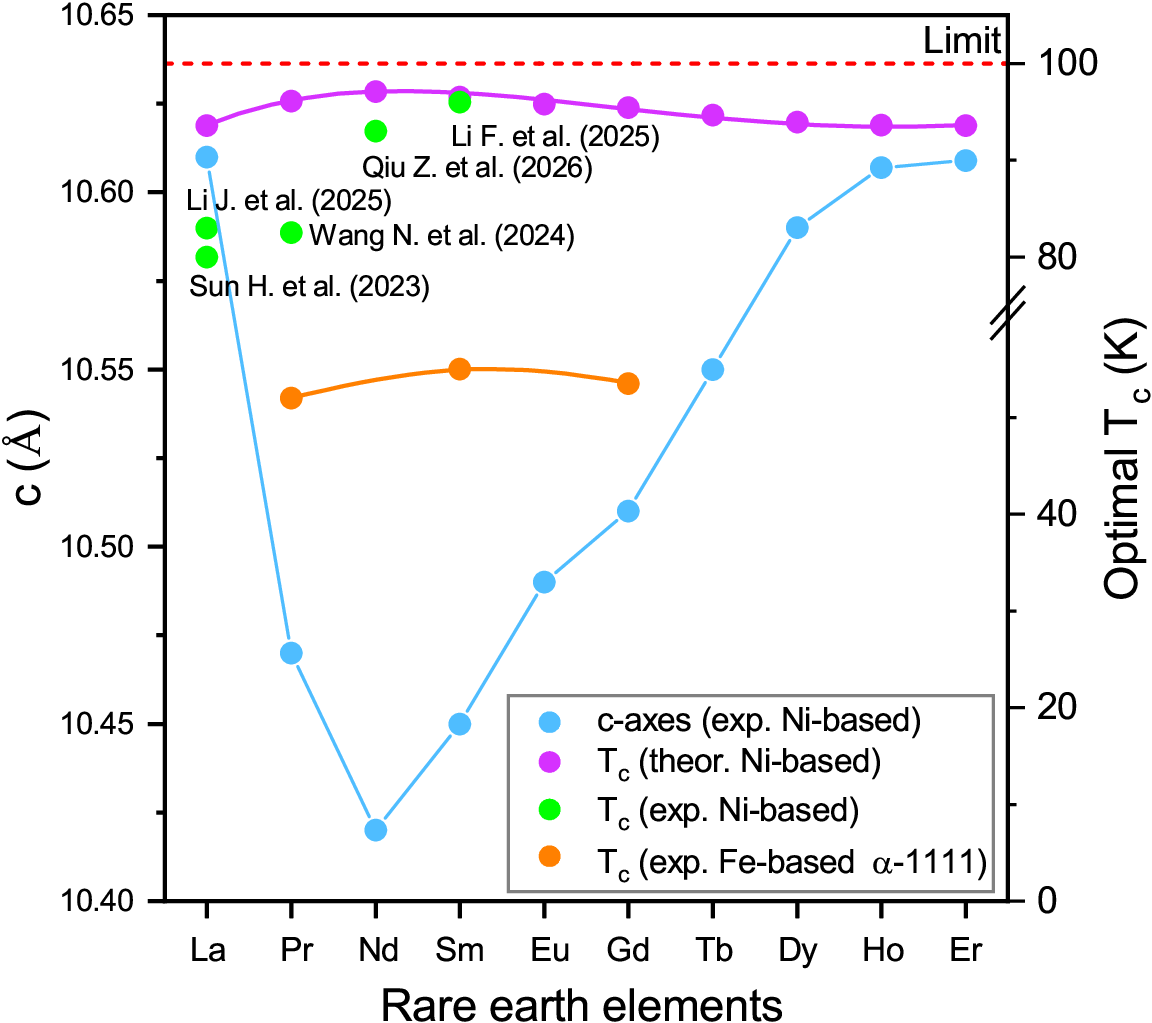}
	\caption{Relationship between Element Substitution and Optimal $T_c$ in Ni-based Superconductors.
		Blue line: experimental $c$-axis lattice parameters of $\mathrm{La_{3-x}R_xNi_2O_{7-\delta}}$ ($\mathrm{R = La}$–$\mathrm{Er}$) from Li \textit{et al.}.
		Purple line: estimated optimal $T_c$ derived from experimental data and Eq. (\ref{dTc}).
		The results indicate that elemental substitution cannot markedly raise $T_c$; existing Ni-based superconductors possess an upper $T_c$ limit of approximately 100 K at $\xi\approx3.54\ \text{\AA}$.
		Orange line: experimental substitution–$T_c$ data of Fe-based superconductors, verifying the limited efficacy of substitution for $T_c$ enhancement. 	}
	\label{figure5}
\end{figure}

The monoclinic-to-tetragonal structural phase transition induces a $\sim2\%$ contraction of $\xi$, 
originating from the more ordered and compact crystal structure. 
According to Eq.~(\ref{dTc}), the superconducting transition temperature change $\Delta T_c\approx3.7\ \text{K}$ 
and the corresponding $T_c=93.4+3.7=97.1\ \text{K}$ are in excellent agreement with the experimentally measured values of $4\ \text{K}$ and $96\ \text{K}$, respectively.

Can elemental substitution further enhance superconducting transition temperature $T_c$? Li et al. studied $\mathrm{La_{3-x}R_xNi_2O_{7-\delta}}$ ($\mathrm{R}$: La–Er rare earths) \cite{Ni96K}, with Fig.~\ref{figure5} blue data showing $c$-axis parameters at maximum $x$. Assuming $c\propto\xi$ (polyhedral quantum well parameter in our theory), optimal $T_c$ of doped samples was estimated using Sm-doped ones as reference. 
Nd doping slightly increases $T_c$ by ~0.5 K (Eq.~(\ref{dTc})). Wang et al. confirmed this limited effect via La$^{3+}$ ($R=1.032$\,\AA) substitution with smaller Pr$^{3+}$ ($R=0.99$\,\AA): internal pressure induces lattice contraction and $\xi$ reduction \cite{chengjg_nature}, yielding $\mathrm{La_{2}PrNi_{2}O_{7-\delta}}$ with $T_c=82.5\ \text{K}$ (~2.5 K higher than $\mathrm{La_3Ni_2O_{7-\delta}}$), consistent with $\Delta T_c\simeq1.85\ \text{K}$ from Eq.~(\ref{dTc}). Recently, Qiu \textit{et al.} employed small-radius $\mathrm{Nd^{3+}}$ to substitute $\mathrm{La^{3+}}$. Under an applied pressure of $26\ \mathrm{GPa}$, bulk superconductivity with zero resistance at $T_\text{c}=93\ \mathrm{K}$ was achieved, while local superconducting signatures were detected up to $100.5\ \mathrm{K}$ via radio-frequency measurements.

Notably, trilayer cuprates host the highest superconducting $T_\text{c}$ among all cuprate families, with Hg-based trilayer systems reaching $164\ \text{K}$ under high pressure \cite{cu164K}. This remarkable performance motivates the exploration of potential high-$T_\text{c}$ behavior in trilayer nickelates \cite{zhu2024superconductivity}. As summarized in Table~\ref{tab:Tc}, the ultrahigh $T_\text{c}$ of Hg-based trilayer cuprates originates from intrinsic octahedral quantum wells with $\xi\approx3.2\ \text{\AA}$, whose effective size can be further reduced via pressure compression. In contrast, trilayer nickelates exhibit three distinct quantum-well characteristic lengths, namely $\xi=3.767\ \text{\AA}$, $3.698\ \text{\AA}$, and $3.767\ \text{\AA}$ \cite{zhu2024superconductivity}. Such larger and asymmetric $\xi$ values result in a suppressed $T_\text{c}$ relative to bilayer nickelates, with values substantially lower than those of cuprate superconductors. 
Furthermore, the maximum $c$-axis lattice difference of $0.19\ \text{\AA}$ shown in Fig.~\ref{figure5} only induces a $4\ \text{K}$ variation in $T_\text{c}$. This weak correlation is further validated by $\alpha$-phase iron-based superconductors (Table~\ref{tab:Tc}), where Pr-doped, Sm-doped, and Gd-doped samples yield $T_\text{c}=52\ \text{K}$, $55\ \text{K}$, and $53.5\ \text{K}$, respectively. Collectively, conventional modulation strategies, including pressure tuning, chemical doping, and ionic substitution, can only marginally enhance the $T_\text{c}$ of existing nickelate superconductors. The empirical upper limit of nickelate $T_\text{c}$ is approximately $100\ \text{K}$, corresponding to a critical quantum-well parameter $\xi\approx3.54\ \text{\AA}$. Therefore, the design and synthesis of new nickelate materials with smaller $\xi$ are essential for achieving further $T_\text{c}$ breakthroughs.

\section{Quantum Well Symmetry Breaking and Fermi Pocket Reconstruction in Artificial Nickelate Multilayers}
\label{sec:1313}

Atomic-layer epitaxy enables the fabrication of four artificial nickelate superstructures, namely 1212, 2222, 2323, and 1313. These structures share identical nickel–oxide building blocks and differ only in atomic stacking sequences. The 1212, 2222, and 2323 phases exhibit ambient-pressure superconductivity with onset \(T_\text{c}\) values of 50 K, 63 K, and 46 K, respectively \cite{xue2026superstructure}. This breakthrough has advanced nickelate superconductivity research and stimulated extensive studies on superconducting thin-film systems \cite{Huang2026MonolayerTrilayerLa3Ni2O7,Chen20261313La3Ni2O7,Ouyang2025PRB,LaBollita2025La5Ni3O11,Nie2026NatureNickelateSuperstructure}. 
Notably, the 2222 and 1313 phases possess identical stoichiometry but distinct superconducting behaviors: the 2222 phase achieves ambient-pressure superconductivity at 63 K, whereas the 1313 phase remains purely metallic and non-superconductive under ambient conditions. Superconductivity in the 1313 phase only emerges under high pressure above 19 GPa, with a low transition temperature of 3.6 K \cite{Huang2026MonolayerTrilayerLa3Ni2O7}. 
Although density functional theory (DFT) and dynamical mean-field theory (DMFT) have been widely adopted to explain this anomalous difference \cite{Chen20261313La3Ni2O7}, existing theoretical predictions are largely inconsistent. Moreover, conventional numerical calculations with flexible tunable parameters fail to clarify the intrinsic superconducting mechanism.

\begin{figure*}[t]
	\includegraphics[width=2\columnwidth]{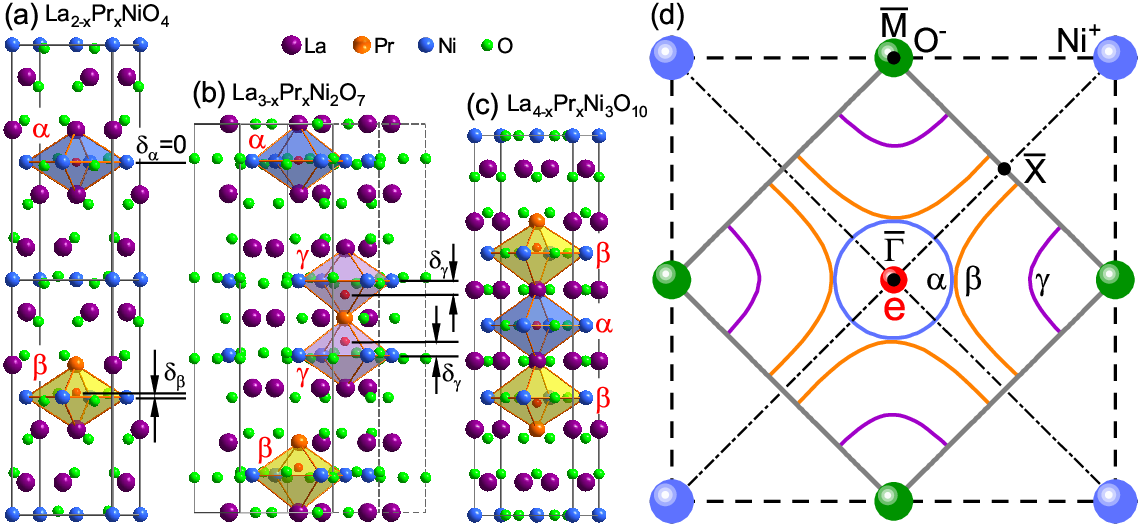}
	\caption{The relationship between quantum wells and Fermi pockets in nickel-based artificial superconducting thin films. (a)--(c) display $\mathrm{La_{2-x}Pr_{x}NiO_{4}}$, $\mathrm{La_{3-x}Pr_{x}Ni_{2}O_{7}}$, and $\mathrm{La_{4-x}Pr_{x}Ni_{3}O_{10}}$, together with their quantum-well structures. The $\gamma$ quantum well present only in (b) enables superconductivity in the 1212, 2222, and 2323 phases; the 1313 phase has no such $\gamma$ quantum well and thus cannot support superconductivity at ambient pressure. (d) shows the correlation between Fermi surface topology and quantum-well electronic states for three quantum-well types. The non-superconducting $\alpha$ quantum well forms nearly circular closed Fermi pockets (blue). Doping-induced symmetry breaking of localized electrons along quantum-well vertices, plus Coulomb attraction from Ni ions, fractures Fermi pockets of $\beta$ and $\gamma$ quantum-well electrons, creating four Fermi arcs with gaps at $\overline{X}$ points. Given $\delta_\gamma\gg\delta_\beta$, Fermi arcs from $\gamma$ quantum wells (purple) possess larger gaps and higher $T_\mathrm{c}$, while $\beta$ quantum wells yield Fermi arcs with smaller gaps (orange) and weaker superconductivity.
	}
	\label{figure6}
\end{figure*}

ARPES measurements confirm that superconductivity is closely associated with the formation of characteristic $\gamma$ Fermi pockets at Brillouin-zone corners \cite{xue2026superstructure}. Here, we systematically investigate the intrinsic differences among four nickelate thin films based on microscopic quantum-well structures. We further clarify the correlation between quantum-well symmetry breaking, Fermi pocket evolution, and superconducting phase formation. Our results demonstrate that high-$T_\text{c}$ superconductivity originates from the bilayer $\text{La}_3\text{Ni}_2\text{O}_7$ subsystem within the 2222 structure, consistent with conclusions from other groups \cite{Ouyang2025PRB,LaBollita2025La5Ni3O11}. The absence of this component fundamentally suppresses Fermi pocket formation and superconductivity.

Figs.~\ref{figure6}(a)--(c) illustrate nickelate superconductors $\mathrm{La_{2-x}Pr_{x}NiO_{4}}$, $\mathrm{La_{3-x}Pr_{x}Ni_{2}O_{7}}$, and $\mathrm{La_{4-x}Pr_{x}Ni_{3}O_{10}}$, corresponding to the 1-, 2-, and 3-unit constituents of the 1212, 2222, 2323 and 1313 series with distinct quantum-well geometries.
Pristine $\mathrm{La_{2-x}Pr_{x}NiO_{4}}$ [Fig.~\ref{figure6}(a)] features mirror-symmetric isolated $\alpha$ quantum wells that localize electrons within Ni planes ($\delta_\alpha=0$), presenting an insulating non-superconducting state. Pr substitution at quantum-well vertices breaks local mirror symmetry to form $\beta$ states, shifting electrons above Ni planes ($\delta_\beta>0$) and inducing weak superconductivity.  
The bilayer phase $\mathrm{La_{3-x}Pr_{x}Ni_{2}O_{7}}$ [Fig.~\ref{figure6}(b)] possesses coupled double quantum wells with three doping-tunable configurations. Its quantum well exhibits hidden mirror symmetry breaking [Fig.~\ref{figure3}(e)], which provides the prerequisite symmetry breaking needed for the superconducting transition. Beyond the conventional $\alpha$ and $\beta$ phases, central-site La substitution generates the asymmetric $\gamma$ phase with strongly broken mirror symmetry ($\delta_\gamma\gg\delta_\beta$), which accounts for high-$T_\mathrm{c}$ superconductivity. 
$\mathrm{La_{4-x}Pr_{x}Ni_{3}O_{10}}$ [Fig.~\ref{figure6}(c)] consists of triple coupled quantum wells. Doping only modifies the upper and lower wells to produce weakly superconducting $\beta$ phases, while the middle well retains the pristine insulating $\alpha$ state. Consequently, the 1313 phase, lacking $\gamma$-type quantum wells, shows no ambient-pressure superconductivity among all synthesized phases \cite{xue2026superstructure,Nie2026NatureNickelateSuperstructure}.

In the quantum-well scenario, the Fermi surface topology originates from photoelectron superposition in localized quantum wells, where each electronic state defines a unique Fermi pocket geometry. As displayed in Fig.~\ref{figure6}(d), electrons in symmetric $\alpha$ wells reach Coulomb force balance, localizing at the NiO plane center at the $\overline{\Gamma}$ point and forming circular Fermi pockets after photoemission.  
Doping-induced symmetry breaking displaces electrons from Ni planes and disrupts force equilibrium. The original symmetric pockets deform into four $\overline{X}$-opened Fermi arcs, whose opening degree positively correlates with the electron displacement $\delta$. The $\beta$ and $\gamma$ states correspond to the orange and purple arcs, respectively. Owing to the much larger $\delta_\gamma$, $\gamma$-phase superconducting arcs concentrate near the Fermi surface $\overline{M}$ corners. The absence of $\gamma$ quantum wells in the 1313 phase eliminates such superconducting Fermi arcs \cite{Nie2026NatureNickelateSuperstructure}.

\section{Conclusion and outlook} \label{sec:conclusion}
This work identifies polyhedral quantum wells as the core structural motif of unconventional superconductors, revealing a universal mechanism for cuprate, iron-based, and nickelate superconductors and clarifying their $T_c$ correlation. Bilayer LSNO shares identical octahedral quantum well structures with YBCO. Based on $T_c=\Lambda/\xi^2$, its monoclinic and tetragonal phases yield predicted $T_c$ values of 93.4 K and 97.1 K, which agree well with experimental measurements. Lanthanide nickelates have an intrinsic $T_c$ limit near 100 K, insensitive to pressure modulation and elemental substitution. Future pursuit of higher-$T_c$ superconductivity should focus on Hg-based cuprate analogs with stronger localization and smaller coherence length $\xi$.

We systematically analyze four synthetic nickelate stacking configurations (2222, 1212, 2323, 1313). Superconductivity originates from bilayer nickelate units, where symmetry breaking of coupled twin quantum wells generates $\gamma$ Fermi pockets and enables ambient-pressure superconductivity. The 1313 phase lacks such quantum-well features and thus cannot sustain ambient-pressure superconductivity. We further demonstrate that Fermi surfaces holographically reflect quantum-well electron states, establishing intrinsic correlations between quantum-well symmetry breaking, Fermi pocket evolution and superconducting performance.

These results verify the predictive and explanatory capacity of QSP theory for the three major superconductor families. Together with its successful description of canonical high-$T_\text{c}$ phenomena \cite{huang1,huang2,huang3}, this work solves the long-standing mechanistic puzzle of high-temperature superconductivity.

\medskip

{\noindent \it Acknowledgements.---} 
The author would like to express special thanks to Prof. WEN Haihu for his invaluable assistance and fruitful discussions. Gratitude is also extended to Prof. ZHANG Junjie and Prof. WANG Ningning for sharing their experimental results.

\medskip

\noindent  \textbf{Data Availability Statement}: All relevant data is explicitly contained within the article.

\noindent \textbf{Competing interests}: The authors declare no competing interests.

%\bibliographystyle{apsrev4-2}

%\bibliographystyle{apsrev4-2-titles} 

%\bibliography{biblio}
%\bibliography{ref}

\end{document}